\documentclass[aps,twocolumn,showpacs,preprintnumbers,amsmath,amssymb,prb]{revtex4-1}
\usepackage{graphicx,subfigure,amsbsy}
\begin{document}

\title{Time-resolved photoemission of correlated electrons driven out of equilibrium}

\author{B. Moritz$^{1,2}$}
\author{T. P. Devereaux$^{1,3}$}
\author{J. K. Freericks$^{4}$}

\address{$^{1}$ Stanford Institute for Materials and Energy Science (SIMES), SLAC National Accelerator Laboratory, Menlo Park, CA 94025, USA}
\address{$^{2}$ Department of Physics and Astrophysics, University of North Dakota, Grand Forks, ND 58202, USA}
\address{$^{3}$ Geballe Laboratory for Advanced Materials, Stanford University, Stanford, CA 94305, USA}
\address{$^{4}$ Department of Physics, Georgetown University, Washington, DC 20057, USA}

\date{\today}

\begin{abstract}
We describe the temporal evolution of the time-resolved photoemission response of the spinless Falicov-Kimball model driven out of equilibrium by 
strong applied fields. The model is one of the few possessing a metal-insulator transition and admitting an exact solution in the time domain. The 
nonequilibrium dynamics, evaluated using an extension of dynamical mean-field theory, show how the driven system differs from two common viewpoints 
--  a quasiequilibrium system at an elevated effective temperature (the ``hot" electron model) or a rapid interaction quench (``melting" of the 
Mott gap) -- due to the rearrangement of electronic states and redistribution of spectral weight. The results demonstrate the inherent trade-off 
between energy and time resolution accompanying the finite width probe pulses, characteristic of those employed in pump-probe time-domain 
experiments, which can be used to focus attention on different aspects of the dynamics near the transition.
\end{abstract}

\pacs{71.10.Fd, 78.47.J-, 79.60.-i, 03.75.-b}

\maketitle

\section{Introduction}
Recently, pump-probe techniques, successfully employed in optical reflectivity studies,~\cite{Cavalieri_reflectance} have been used to extend 
photoemission spectroscopy~\cite{ARPES_Rev} (PES) to the time domain (time-resolved PES or tr-PES) in the femtosecond and attosecond 
regimes.~\cite{Schmitt_PP,Perfetti_PP_1,Perfetti_PP_2,Perfetti_PP_3,Lisowski_PP,Cavalieri_PP}  These advances open the possibility of observing and 
controlling dynamics on time-scales relevant to correlated electronic processes,~\cite{Krausz_atto,Cavalieri_Science,Cavalieri_cond-mat} specifically 
in optical pump-probe radio frequency (rf) cold atom and table-top laser experiments, as well as using free-electron laser facilities, such as the 
Free-electron LASer in Hamburg (FLASH) and the Linac Coherent Light Source (LCLS), to conduct pump-probe extreme ultra-violet or x-ray photoemission 
and scattering studies in the time domain. 

Most pump-probe experiments have been interpreted in terms of ``hot" electrons, effectively equilibrated at highly elevated 
temperatures.~\cite{Perfetti_PP_1,Perfetti_PP_2,Perfetti_PP_3,Freericks_4}  While this approach can capture a redistribution of spectral intensity 
through the change in the Fermi distribution function and thermal rearrangement of electronic states that experiments observe on picosecond time scales, 
it can not account properly for the out-of-equilibrium rearrangement of accessible electronic states nor capture the nonequilibrium redistribution of 
spectral weight that accompanies pump pulses with the high excitation densities needed to drive phase transitions or excite collective modes 
characteristic of correlated electron systems on ultrashort time scales in the femtosecond or attosecond regimes.

As a test case, we have chosen to study the effect of strong driving fields on a simple model system for which the hot electron model definitively 
breaks down and where the effect of the driving field does not mimic an interaction 
quench~\cite{Eckstein_quench_PRL,augsburg,Werner_quench_PRL,Werner_quench} or ``melting" of the Mott gap.  We examine the temporal evolution of the 
tr-PES response for the spinless Falicov-Kimball model at half-filling, driven by a large amplitude, dc electric field toward a nonequilibrium 
steady-state.~\cite{Freericks_1,Freericks_2}  The Falicov-Kimball model is one of the simplest correlated electron models and it has a Mott-Hubbard 
metal-insulator transition (MIT) at half-filling; to this point, it is the only model for which an exact nonequilibrium impurity solution has been 
developed in time-dependent fields with a time range long enough to evaluate tr-PES.  In particular, the temperature invariance of the equilibrium 
density of states (DOS) for this model makes a comparison to hot electrons at long time delays relatively straightforward.  We find that the 
spectral intensity develops regular Bloch-like oscillations for weak metallic correlations that become sharply damped approaching and crossing the 
MIT.  The results elucidate the out-of-equilibrium behavior of a simple correlated electron system observed using tr-PES as a probe.

\section{Method}
We determine the real time dynamics for the model on the hypercubic lattice in infinite dimensions ($d=\infty$)  using nonequilibrium dynamical 
mean-field theory (DMFT).~\cite{Freericks_1,Freericks_2,augsburg}  This method yields the double-time contour-order Green's function (GF) 
$G^{\mathcal{C}}(t,t')$ within the Kadanoff-Baym-Keldysh formalism.~\cite{Keldysh,Kadanoff-Baym}  The system begins in thermal equilibrium at time 
$t=t_{\rm min}$ and temperature $T$ before an electric field, applied at $t=0$, breaks time-translation invariance.  The system evolves under the 
influence of this field to a maximal time $t_{\rm max}$.  This defines the Keldysh contour $\mathcal{C}$ used in the formalism.  The contour-ordered 
GF encodes both the retarded GF, determining the equilibrium as well as nonequilibrium arrangement of states, and the lesser GF, specifying the 
distribution of electrons among these states, together with other physically relevant GFs.  In particular, the lesser GF, related to the PES 
response, is given by the Keldysh contour-ordered quantity,
\[
G_{ij}^{<}(t,t') = i{\rm Tr}\left[\exp^{-\beta H_{\rm eq}}\,c^{\dag}_{j}(t')c^{}_{i}(t)/Z_{\rm eq}\right],
\]
in the Heisenberg picture, where $t$ lies on the upper real time branch and $t'$ on the lower real time branch of the Keldysh contour and ``${\rm 
eq}$" denotes equilibrium $(t<0)$ quantities at the initial temperature $T$.

The equilibrium Hamiltonian is given by
\[
H_{\rm eq} = -\frac{t^{*}}{2\sqrt{d}}\sum_{<ij>}(c^{\dag}_{i}c_{j} + h.c.) - \mu\sum_{i}c^{\dag}_{i}c_{i} + U\sum_{i}w_{i}\,c^{\dag}_{i}c_{i},
\]
describing the hopping, $t^{*}$, of conduction electrons between lattice sites with a filling controlled by the chemical potential, $\mu$, that 
experience an on-site interaction, $U$, with another species of localized electrons with an occupation $w_{i}$.  $U_{c}=\sqrt{2}\,t^{*}$ is the 
critical interaction strength for the MIT at half-filling.  Throughout this work, the energy unit is taken to be $t^{*}$.  As an aid to the reader in 
understanding the relevant time and field scales for this paper,  consider those set by the hopping integral $t^{*} = 250$ meV and hypercubic lattice 
spacing $a = 3$ \AA; the corresponding unit of time is $\sim 16$ fs and that for the dc driving field $E$ is $\sim 13$ mV/\AA.  The nonequilibrium 
DMFT formalism proceeds in essentially the same manner as the iterative approach applied in equilibrium~\cite{Jarrell_DMFT} where all quantities now 
have two time indices.~\footnote{Quadratic extrapolation of the Green's functions to zero step size on the discrete Keldysh contour ensures that sum 
rules for the spectral moments~\cite{sumrules} are satisfied within a few percent even for strong correlations and high fields.}

The driving term is modeled by a spatially uniform constant dc electric field along the $(1, 1, 1,\ldots)$ hypercubic body diagonal high symmetry 
direction to simplify evaluation of the noninteracting GF.~\cite{Freericks_2}  The spatially uniform vector potential, in a gauge with zero scalar 
potential (Hamiltonian gauge), associated with this driving field varies linearly in time and enters through a Peierls' 
substitution.~\cite{Peierls_sub}  

We determine the real frequency spectral intensity using a finite width probe pulse that samples the real time dynamics of the driven nonequilibrium 
system. The probe pulse envelope in a tr-PES experiment can be well approximated by a Gaussian waveform
\[
s(t) = \frac{1}{\sigma\sqrt{\pi}}\exp^{-(t-t_{o})^{2}/\sigma^2},
\]
where $t_{o}$ measures the time delay with respect to the application of the driving field and $\sigma$ measures its effective temporal width.  The 
tr-PES response function is then a probe pulse weighted relative time Fourier transform of the lesser GF.~\cite{Freericks_3,Freericks_4,augsburg}

\section{Results and Discussion}
\begin{figure}[t]
\includegraphics[width=\columnwidth]{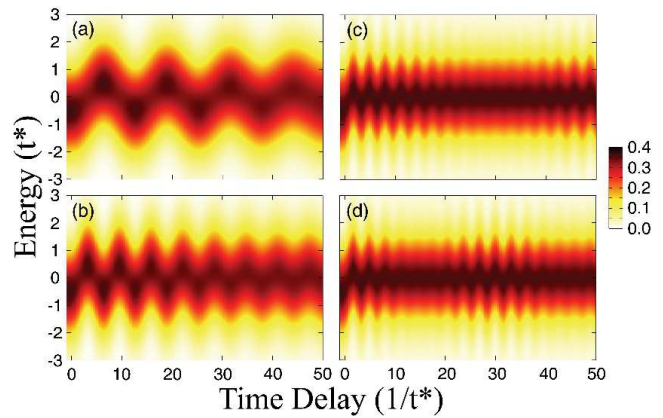} \caption{(Color online.) Tr-PES intensity (falsecolor or grayscale) as a function of photoelectron 
energy and time delay for weakly correlated metallic systems $U<U_{c}$ [(a) - (c) $U=0.125$ and (d) $U=0.25$].  A probe pulse of characteristic width 
$\sigma=1.0$ samples the nonequilibrium dynamics of the system driven by fields of strength (a) $E=0.5$, (b) $E=1.0$, and [(c) and (d)] $E=2.0$.  Bloch 
oscillations with a period proportional to $1/E$ develop almost immediately following application of the field at $t=0$ and an additional amplitude modulation 
with a period proportional to $1/U$ appears for the strongest driving fields [(c) and (d)].\label{fig:1}}
\end{figure}

Results are shown in Fig.~\ref{fig:1} for metallic systems with weak correlations $U<U_{c}$ [(a) - (c) $U=0.125$ and (d) $U=0.25$] driven by applied 
fields with different strengths $E$. Each panel shows the spectral intensity (falsecolor or grayscale), as a function of binding energy and time 
delay (obtained from data generated with a discretization in time equal to $0.1$).  For the lowest field strengths [Fig.~\ref{fig:1}(a), $E=0.5$, and 
Fig.~\ref{fig:1}(b), $E=1.0$] the spectral intensity develops regular Bloch oscillations, with a period proportional to $1/E$, damped by 
correlations.  For sufficiently large fields [Fig.~\ref{fig:1}(c), $U=0.125$, and Fig.~\ref{fig:1}(d), $U=0.25$, with $E=2.0$], an additional amplitude mode, 
characterized by ``beats" in the spectral intensity, appears with a period proportional to $1/U$.

\begin{figure}[t]
\includegraphics[width=\columnwidth]{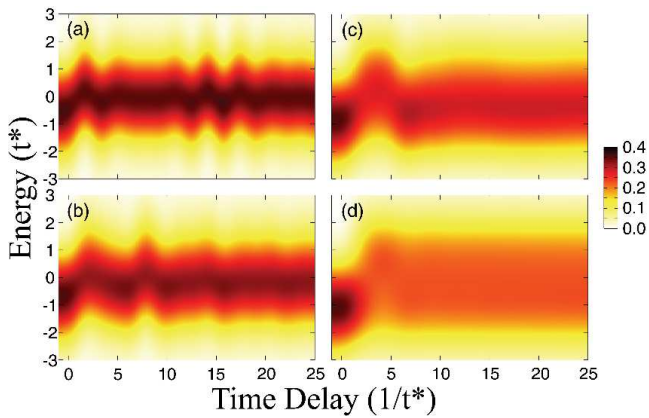}
\caption{(Color online.) Tr-PES intensity (falsecolor or grayscale) for various correlations in both metallic $U<U_{c}$ [(a) $U=0.5$ and (b) $U=1.0$] 
and insulating $U>U_{c}$ [(c) $U=1.5$ and (d) $U=2.0$] regimes driven by fields of strength $E=2.0$ and sampled using a probe pulse of width 
$\sigma=1.0$.  Bloch oscillations become increasingly damped approaching and crossing the MIT.\label{fig:2}}
\end{figure}

Figure~\ref{fig:2} shows results for interaction strengths approaching and crossing the MIT.  Oscillations associated with the driving field are 
still apparent for metallic systems $U<U_{c}$ [Fig.~\ref{fig:2}(a), $U=0.5$, and Fig.~\ref{fig:2}(b), $U=1.0$], but the amplitude mode and increased damping 
lead to a rather irregular temporal evolution.  For $U=1.5$, just above $U_{c}$, the damping is severe enough to suppress oscillations for all but the 
shortest time delays.  Further increase in the interaction strength exacerbates these effects.  The behavior of these oscillations is the tr-PES 
analog of that found for the instantaneous current response, evaluated from the \emph{equal-time} lesser GF.~\cite{Freericks_2}

The results shown in Figs.~\ref{fig:2}(c) and ~\ref{fig:2}(d) naively suggest, at least for $U>U_{c}$, that the dc field drives the system toward a 
nonequilibrium steady-state in which the Mott gap has melted or the interaction strength has been quenched to a smaller value $U<U_{c}$, resulting 
in significant spectral weight at and above the equilibrium Fermi level (Energy = $0$) for long time delays.  However, these observations are 
merely artifacts of the trade-off between energy and time resolution associated with the relatively narrow Gaussian probe pulses used in 
Fig.~\ref{fig:2} to highlight the temporal dynamics.  In the transient response regime there is no time translation invariance and the probe width 
affects {\it both} the temporal and energy resolution.  However, the conventional Fourier uncertainty relations would be recovered in the 
steady-state regime at long times.~\cite{Freericks_3,augsburg}

\begin{figure}[t]
\includegraphics[width=\columnwidth]{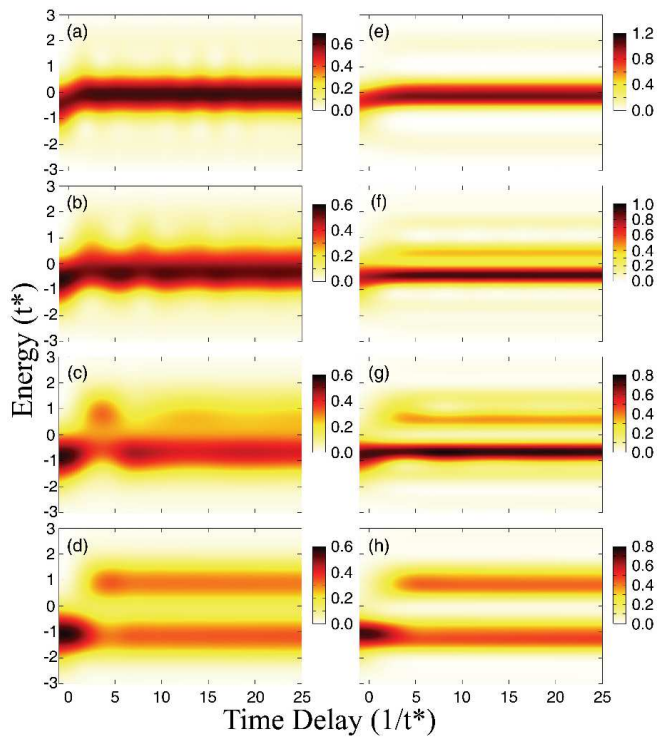}
\caption{(Color online.) Tr-PES intensities (falsecolor or grayscale) from the metallic to insulating regimes for different probe pulse widths.  
Together with increased energy resolution, wider probe pulses lead to qualitative changes in the temporal evolution of the PES response and to a 
sharpening of spectral features as a function of energy.  The parameters are [(a) and (e)] $U=0.5$, [(b) and (f)] $U=1.0$, [(c) and (g)] $U=1.5$, and 
[(d) and (h)] $U=2.0$ for $\sigma=2.0$ and $4.0$, respectively, all driven by fields of strength $E=2.0$.\label{fig:3}}
\end{figure} 

Figure~\ref{fig:3} shows results similar to those of Fig.~\ref{fig:2}, but for wider probe pulses.  Note that the increase in width leads to a 
suppression of temporal dynamics, except at the shortest time delays, and sharper spectral features as a function of energy.  For the weakly 
correlated metal [Figs.~\ref{fig:3}(a) and ~\ref{fig:3}(e), $U=0.5$], the increased width suppresses the regular Bloch oscillations.  The spectral intensity 
approaches the steady-state at long times including the redistribution of weight into faint high- and low-energy satellites at $\sim \pm E$ associated 
with the rungs of the Wannier-Stark (WS) ladder.~\cite{Wannier_Stark}  The WS ladder describes the rearrangement of electronic states into periodic 
resonances in energy within a conventional metal or weakly interacting system due to the application of a strong driving field.  The spectral 
intensity for the strongly correlated metal [Figs.~\ref{fig:3}(b) and ~\ref{fig:3}(f), $U=1.0$] behaves similarly with the appearance of additional WS satellites 
and a suppression of weight near zero energy attributable to the rearrangement of accessible electronic states~\cite{Freericks_2} and not an \emph{ad 
hoc} change to the interaction strength.

For $U \sim U_{c}$ [Figs.~\ref{fig:3}~(c) and ~\ref{fig:3}~(g)] there is still a significant redistribution of weight across the equilibrium Mott gap at short 
time delays before relaxing and partially recovering at longer times.  Further increase in the probe pulse width ($\sigma=4.0$) suppresses the 
temporal evolution but does reveal a modified real frequency structure.  Figures~\ref{fig:3}~(d) and ~\ref{fig:3}~(h) show results for $U=2.0$.  In this case, 
spectral weight is distributed between two main features and the temporal evolution appears only through a modulation of the spectral intensity 
within these features.  The ability to track changes in the spectral intensity persists only for the shortest time delays after applying the 
driving field and the spectra quickly approach that characteristic of the nonequilibrium steady-state.  In each of the cases presented in 
Fig.~\ref{fig:3}, the tr-PES response clearly is not indicative of an interaction quench with subsequent melting of the equilibrium Mott gap. 

\begin{figure}[t]
\includegraphics[width=\columnwidth]{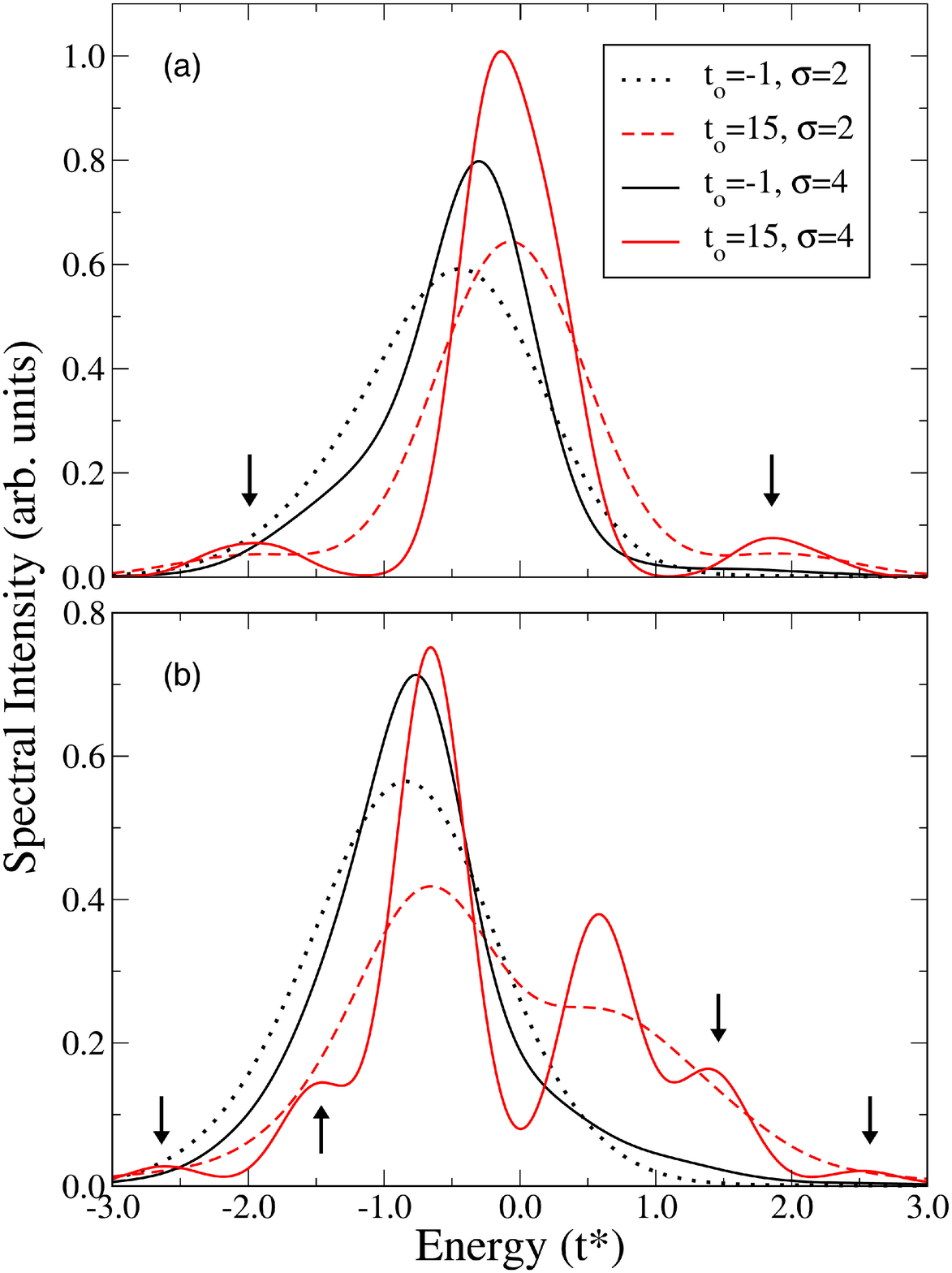}
\caption{(Color online.) Equilibrium and nonequilibrium steady-state response for (a) $U=0.5$ and (b) $U=1.5$ with a driving field strength $E=2$.  
These results correspond to time delay cuts at $t_{o}=-1$ and $t_{o}=15$ from the results shown in Figs.~\ref{fig:3}(a),~\ref{fig:3}(c),~\ref{fig:3}(e), and 
~\ref{fig:3}(g), sampled with two different probe pulses of width $\sigma=2.0$ and $\sigma=4.0$, respectively. The nonequilibrium steady-state response 
(time delay $t_{o}=15$) shows high and low energy Wannier-Stark satellites (highlighted by arrows) and overall line shape, including a sharpening of spectral 
features, incompatible with both the simple hot electron view and melting of the Mott gap or an interaction quench with applied field.\label{fig:4}}
\end{figure}

Finally, we compare the response in equilibrium (taken at time delay $t_{o} = -1$) to the response approaching the regime of the nonequilibrium 
steady-state (taken at time delay $t_{o} = 15$).  Figure~\ref{fig:4} shows this comparison for both a metallic [Fig.~\ref{fig:4}(a), $U=0.5$] and 
insulating [Fig.~\ref{fig:4}(b), $U=1.5$] system probed by pulses of widths $\sigma=2$ and $4$.  The response in equilibrium (black curves) 
essentially matches the equilibrium DOS multiplied by the Fermi distribution function and convolved with an energy resolution function accounting for 
the finite temporal width of the probe pulse.  The energy resolution improves with wider probe pulses, manifest in a sharpening of spectral features 
between $\sigma=2$ and $4$.

The response in the nonequilibrium steady-state (red or gray curves) shows qualitative differences to those in equilibrium.  Note that the 
equilibrium DOS in the Falicov-Kimball model at half-filling is symmetric with respect to the Fermi level and temperature independent; therefore, at 
a higher effective temperature, the spectral weight should be redistributed to at most one higher-energy feature above the equilibrium Fermi level.  
However, in these cases both high and low energy WS satellites are found in the response, indicated by arrows, and the features are far sharper than 
those in equilibrium.  For the insulator shown in Fig.~\ref{fig:4}(b), there is even a suppression of spectral weight at the Fermi level.  The 
satellites are more pronounced for weak correlations and wider probe pulses [see Fig.~\ref{fig:3} and compare Figs.~\ref{fig:4}(a) and ~\ref{fig:4}(b)]; however, 
they are present in the response for all cases, highlighted here by systems on both sides of the MIT.  Overall, this behavior precludes a simple 
quasiequilibrium description of the out-of-equilibrium response of the system in terms of an elevated effective temperature -- the ``hot" electron 
model.

\section{Conclusions}
The current model captures the formation of damped Bloch oscillations for weakly correlated metals.  The oscillations, typically suppressed in real 
materials due to scattering from phonons and impurities, not included in this model, are simply damped here by electron correlations.  Experiments 
corresponding to the conditions represented in these simple model calculations potentially could be performed in ultracold atomic systems by 
generalizing equilibrium rf techniques~\cite{jin_rf} to nonequilibrium situations.  The experiment would involve mixtures of light fermions with heavy
fermions or heavy bosons at low temperature (Li$^6$-K$^{40}$ mixtures for the former or Li$^6$-Cs$^{133}$ or K$^{40}$-Rb$^{87}$ mixtures for
the latter), applying a driving field to generate the Bloch oscillations.  This ``field'' could be gravity or a detuning of the counter-propagating lasers 
which ``pulls'' the optical lattice through the atomic clouds. The rf pulses would need to have a duration which is short enough in time to observe 
the Bloch oscillations in the time domain after a time-of-flight image.  This type of experiment may be cleaner than those performed on 
conventional condensed matter systems because the driving fields will not interfere with the time-of-flight detection used to probe these systems. 
For conventional condensed matter systems, using ultrafast probe pulses to determine the tr-PES response of a system driven by a strong electric 
field toward a nonequilibrium steady-state may be challenging to replicate experimentally.  The ability to observe short-time behavior (on the scale 
of femtoseconds) opens the possibility of observing oscillations before they become damped (something that may occur at picosecond time scales, 
especially in weakly correlated materials). It is conceivable that the WS ladder or Bloch oscillations could be seen within the duration of a wider 
pump excitation using a particularly sharp probe pulse.  This could be true for FEL sources with exceptionally large amplitude pump pulses.  
However, this method would require a modification of existing synchronization techniques and improvements in probe temporal resolution to reach the 
necessary time scales.

\acknowledgments The authors would like to thank D. Jin, P. S. Kirchmann, H. R. Krishnamurthy, F. Schmitt, and M. Wolf for valuable discussions.  
B.M. and T.P.D. were supported by the U.S. Department of Energy (DOE), Office of Basic Energy Sciences (BES), under Contract No. DE-AC02-76SF00515 (SIMES and 
Single Investigator and Small Group Research grant).  J.K.F. was supported by the National Science Foundation under Grant No. DMR-0705266 for the 
generation of the nonequilibrium data and from the U.S. DOE, BES, under Grant No. DE-FG02-08ER46542 for the analysis of the data. The collaboration was 
supported by the Computational Materials Science Network program of the U.S. DOE, BES, Division of Materials Science and Engineering, under Grant No. 
DE-FG02-08ER46540. The data analysis was made possible by the resources of the National Energy Research Scientific Computing Center (via an Innovative and 
Novel Computational Impact on Theory and Experiment grant) which is supported by the U.S. DOE, Office of Science, under Contract No. DE-AC02-05CH11231.

\bibliography{Moritz_tr-ARPES}

\end{document}